# Local and Global Superconductivity in Bismuth

Luis A. Baring, Robson R. da Silva, and Yakov Kopelevich

Instituto de Física "Gleb Wataghin", Universidade Estadual de Campinas, Unicamp

13083-859, Campinas, São Paulo, Brasil

ABSTRACT

We performed magnetization $M(H,T)$ and magnetoresistance $R(T,H)$ measurements on powdered (grain size ~ 149 μm) as well as highly oriented rhombohedral (A7) bismuth (Bi) samples consisting of single crystalline blocks of size ~ 1x1 mm$^2$ in the plane perpendicular to the trigonal c-axis. The obtained results revealed the occurrence of (1) local superconductivity in powdered samples with $T_c(0)$ = 8.75 ± 0.05 K, and (2) global superconductivity at $T_c(0)$ = 7.3 ± 0.1 K in polycrystalline Bi triggered by low-resistance Ohmic contacts with silver (Ag) normal metal. The results provide evidence that the superconductivity in Bi is localized in a tiny volume fraction, probably at intergrain or Ag/Bi interfaces. On the other hand, the occurrence of global superconductivity observed for polycrystalline Bi can be accounted for by enhancement of the superconducting order parameter phase stiffness induced by the normal metal contacts, the scenario proposed in the context of "pseudogap regime" in cuprates [E. Berg et al., PRB **78**, 094509 (2008)].

In spite of semimetallic bismuth (Bi) has been extensively studied for decades, it still possesses a number of puzzling physical properties that include, for example: (i) a huge non-saturating magnetoresistance [1-3], magnetic-field-driven metal-insulator-type transition(s) [4], strong diamagnetism [5, 6], possible fractional quantum Hall effect [7], and superconductivity [8-14].

In the present work we focus our attention on superconducting properties of Bi. While single crystalline rhombohedral (A7) Bi is not superconducting, at least down to 50 mK, the superconductivity has been observed for high pressure phases: Bi-II (monoclinic, p = 2.55 GPa), Bi-III (tetragonal, p = 2.7 GPa), and Bi-V (bcc, p = 7.7 GPa), having the superconducting transition temperature $T_c$ = 3.9 K, 7.2 K and 8.3 K, respectively (see e. g. Ref. [15]).

On the other hand, the surface or interface superconductivity has been found in granular films [10, 11] and nanowires [12] made of Bi clusters with rhombohedral structure, bicrystals [13], and amorphous Bi [16]. One possible explanation for the superconductivity occurrence is the structural reconstruction at the grain or bicrystal interfaces. If such a reconstruction takes place indeed, the observation of superconductivity with $T_c$ = 21 K [14] suggests that the phase(s) are different from that obtained under pressure. The strong sensitivity of $T_c$ to adsorbed gases [10] also suggests that the origin of the superconductivity may be different. Besides, earlier experiments [8] demonstrated that the ohmic contact between single crystalline Bi and metal, e. g. silver (Ag), may trigger the superconductivity.

Aiming to shed more light on the origin of the surface/interface superconductivity in Bi, in this work we performed magnetic and magnetotransport measurements on highly oriented polycrystalline rhombohedral (A7) Bi as well as rhombohedral Bi powder with the grain size of ~ 149 μm – mesh -100 – (Aldrich, 99.999% pure the impurity content: Si – 4 ppm, Cu – 2 ppm, and Fe – 1 ppm). Magnetization M(T, H) measurements were carried out with a commercial superconducting quantum interference device magnetometer (Quantum Design) MPMS5. Low-frequency (f = 1 Hz) and dc magnetoresistance measurements R(T,B) were performed by means of PPMS (Quantum Design) and Janis 9T-magnet He$^4$ cryostats using standard four (or two) - probe methods with Ag electrodes placed on the sample surface.

Figure 1 presents temperature dependencies of normalized magnetization $M_{ZFC}(T)/|M_{ZFC}(60\ K)|$ obtained for Bi powder in the zero-field-cooling (ZFC) regime for various applied magnetic fields. The results provide a clear experimental evidence for the superconducting transition that takes place at T ~ 8.7 K. The right inset gives both $M_{ZFC}(T)/|M_{ZFC}(60\ K)|$ and $M_{FCC}(T)/|M_{FCC}(60\ K)|$ obtained for H = 100 Oe, where $M_{FCC}(T)$ is the magnetization measured in the field cooled on cooling (FCC) regime. In the FCC regime the superconducting transition is not visible indicating an immeasurably small Meissner fraction being consistent with the surface/interface superconductivity or/and strong vortex pinning (see below). The superconducting shielding fraction is estimated to be ~30 ppm. The resistance measurements (the left inset in Fig. 1) revealed no signature for the superconducting transition, implying that the superconductivity is localized within some sort of decoupled superconducting grains or islands large enough to carry vortices. Noting that our X-ray (Θ-2Θ geometry) analysis revealed only rhombohedral A7 Bi and 2% of $Bi_2O_3$, which is not superconducting.

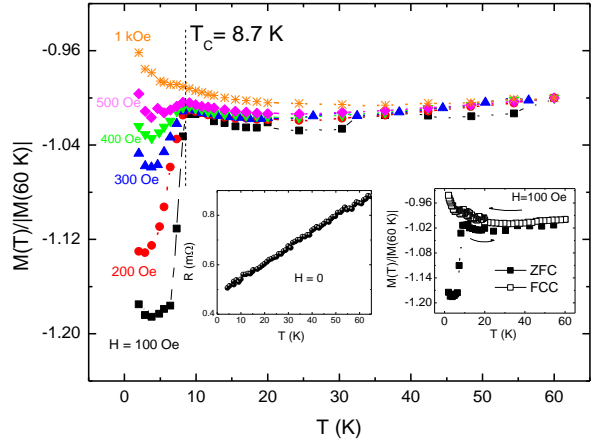

Fig.1. Normalized magnetization vs. temperature measured for Bi powder at various magnetic fields in the ZFC regimes, showing the occurrence of superconducting transition at $T_c$ = 8.7 K. The right inset demonstrates the invisibility of the Meissner fraction verified in field cooled on cooling (FCC) measurements. The left inset gives the temperature dependence of the resistance R(T).

To characterize further the superconducting phase, we performed M vs. H measurements at various temperatures. The low-field portions of $M_{ZFC}(H)$ isotherms obtained after subtraction of irrelevant here orbital diamagnetic signal are shown in Fig. 2. The measured magnetization hysteresis loop, see Fig. 3, provides the unambiguous evidence that our sample is a type-II superconductor with a strong vortex pinning [17]. The results of Fig. 2 allows one to determine the lower critical field $H_{c1}(T)$, in which, the magnetization values were multiplied by factors (in brackets) to better visualize the data. For $H < H_{c1}(T)$, the $|M_{ZFC}(T)|$ linearly increases with field, see Fig. 2, as in the Meissner phase, and it deviates from the straight line when the field starts to penetrate the superconducting regions in the form of vortices. The obtained $H_{c1}(T)$ is shown in Fig. 4 that can be best fitted by the two-fluid-model equation $H_{c1}(T) = H_{c1}(0)[1 - (T/T_c)^4]$ with $H_{c1}(0) = 129 \pm 1$ Oe, and $T_c = 8.75 \pm 0.05$ K.

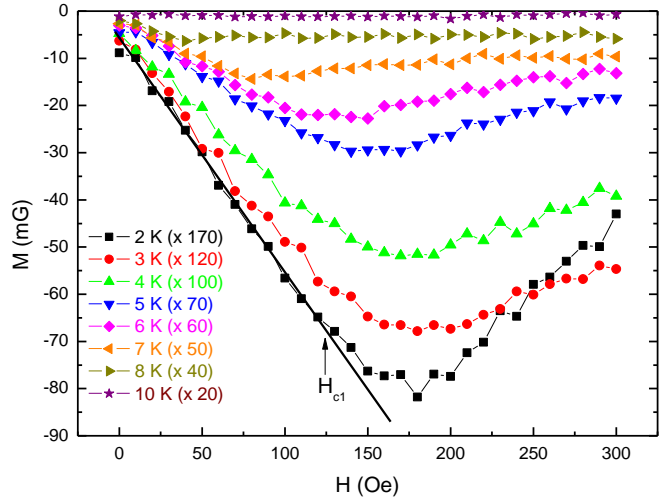

Fig. 2. The low-field portions of M(H) isotherms obtained in ZFC regime after subtraction of the here irrelevant orbital diamagnetic signal. The magnetization values were multiplied by factors (in brackets) to better visualize the data. The lower critical field $H_{c1}(T)$ was defined, as exemplified for M(H) obtained at $T = 2$ K, by the field of departure from the Meissner behavior, represented by the straight line.

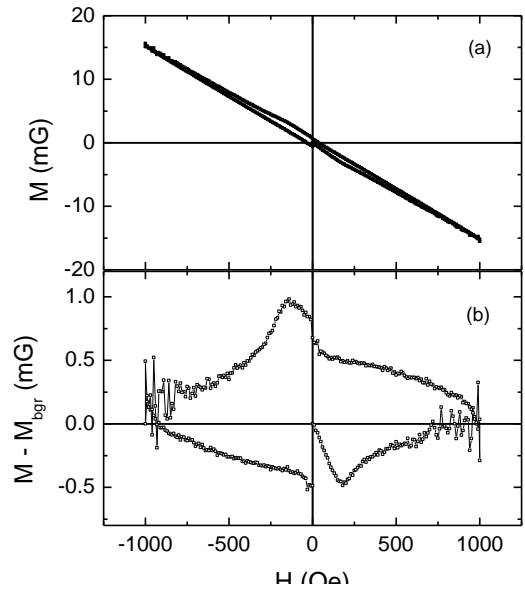

Fig. 3. (a) Magnetization hysteresis loop M(H) measured at $T = 2$ K; (b) the same M(H) after subtraction of the diamagnetic background magnetization $M = -\chi H$, where $|\chi| = 0.0151$ mG/Oe. The results demonstrate that the measured Bi powder is a type-II superconductor with a strong vortex pinning [17].

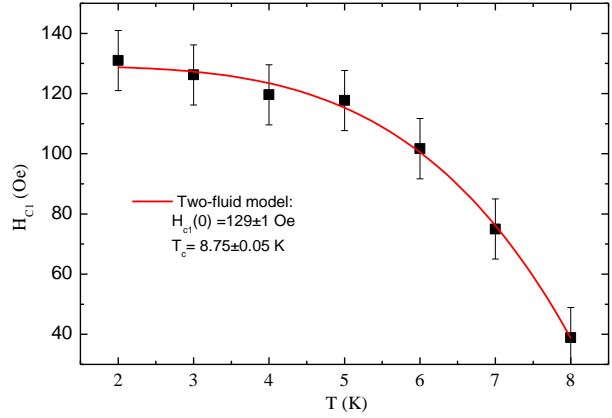

Fig. 4. . Lower critical field $H_{c1}(T)$ obtained from the data of Fig. 2. The red line corresponds to the two-fluid-model equation $H_{c1}(T) = H_{c1}(0)[1 - (T/T_c)^4]$ with $H_{c1}(0) = 129\pm1$ Oe, and $T_c = 8.75 \pm 0.05$ K.

Figure 5 illustrates that the superconductivity vanishes after the sample annealing, pointing out on the surface character of the phenomenon. Much smaller superconducting signal observed in Bi polycrystalline samples [4] consisting of single crystalline blocks of size ~ 1x1 mm$^2$ in the plane perpendicular to the trigonal c-axis [4] (see also Fig. 6) provides another piece of evidence that the enhanced surface area in powdered samples plays an important role in the superconductivity occurrence.

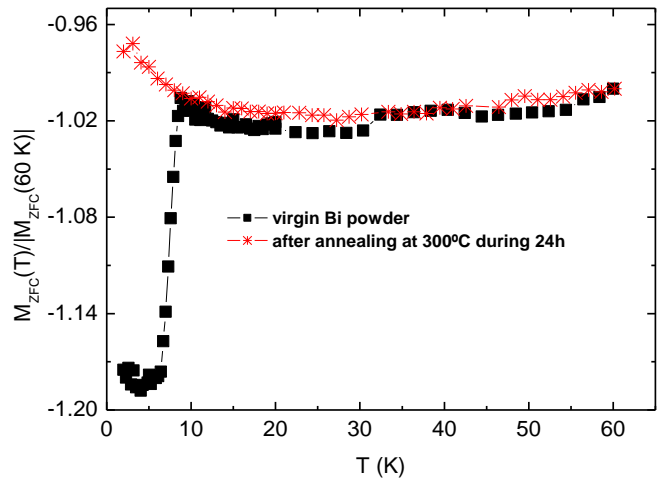

Fig. 5. Reduced ZFC magnetization measured for virgin Bi powder and after the sample annealing at T = 300 ºC during 24 h in Ar atmosphere.

In what follows, we show that both local and global (zero resistance) superconductivity can be induced in Bi contacted by a normal metal.

We performed magnetoresistance measurements on highly oriented rhombohedral polycrystalline bismuth samples (Fig. 6), with magnetic field applied parallel to the trigonal c-axis. Several Bi samples obtained from the same bar were used in transport measurements. Four non-superconducting contacts (Ag) in the standard configuration were placed on the sample surface, as exemplified by Fig. 7 for one of the studied samples. In four-probe measurements, $R_{23} = V_{23}/I_{14} \equiv R_1$ where the current $I_{14}$ flows between contacts 1 and 4 and the voltage is recorded between contacts 2 and 3. In two-probe measurements, the resistance $R_{14} = V_{14}/I_{14} \equiv R_2$ is obtained applying the current $I_{14}$ between contacts 1 and 4 and measuring the voltage $V_{14}$ at the same contacts.

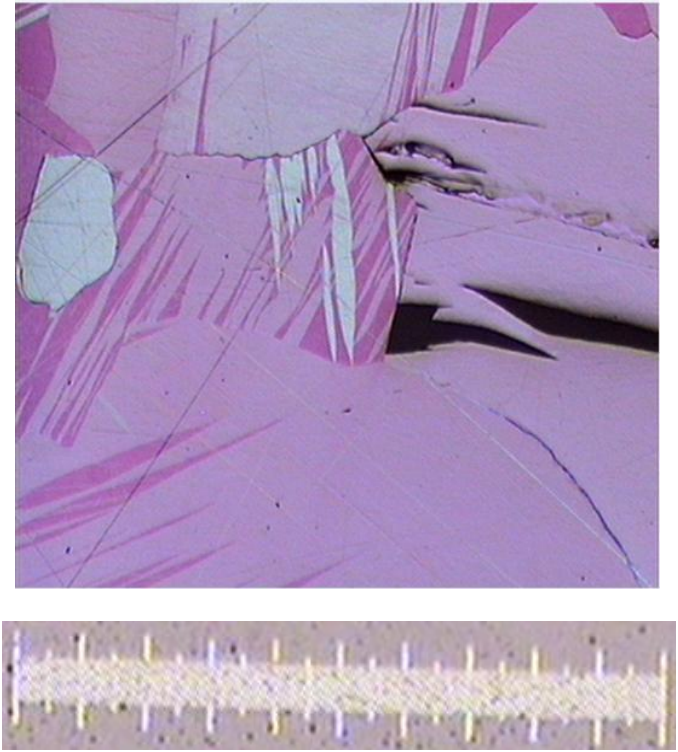

1 mm

Fig. 6. False color optical image of Bi polycrystalline sample studied in Ref. [4] and the present work.

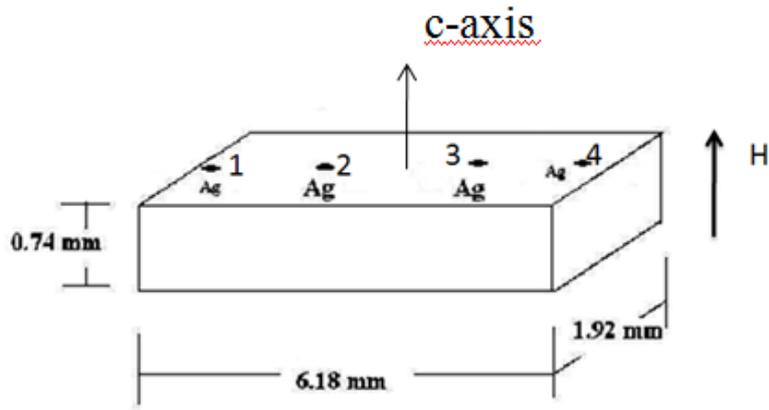

Fig.7. Four non-superconducting contacts (Ag) placed on the surface of Bi polycrystalline sample consisting of single crystalline blocks of size ~ 1 x 1 mm$^2$ (Fig. 6) in the plane perpendicular to the trigonal c-axis.

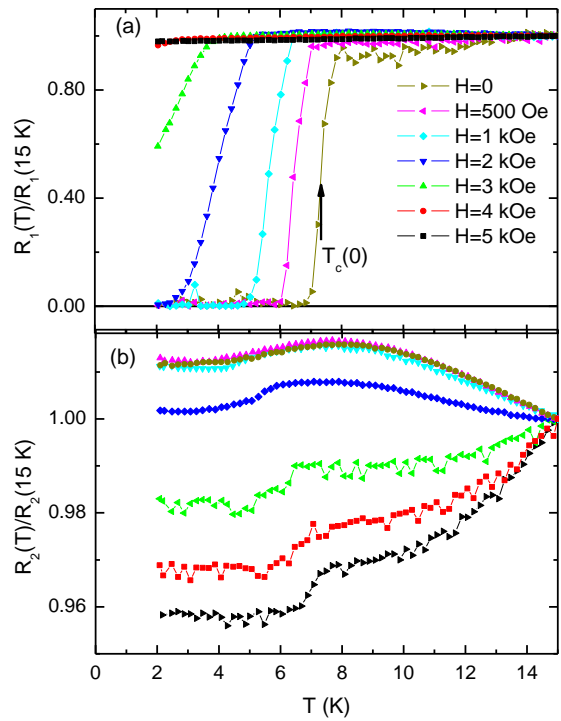

Fig.8. Reduced resistance measured for bulk polycrystalline Bi sample (Fig. 7) at zero and various applied magnetic fields; (a) four-probe $R_1 = V_{23}/I_{14}$ and (b) two-probe $R_2 = V_{14}/I_{14}$ measurements. Arrow in (a) marks the zero-field superconducting transition temperature $T_c(0) = 7.3 \pm 0.1$ K determined at the maximum of the derivative $dR_1(T)/dT$.

Figures 8(a) and 8(b) present results of four- and two-probe resistance measurements, respectively. The salient feature of the data given in Fig. 8(a) is the zero-resistance state that takes place below the superconducting transition temperature $T_c(H)$. The superconductivity is also evident from the resistance drop measured in two-probe configuration (Fig. 8, b).

Based on these data, we got the upper critical field $H_{c2}(T)$, as shown in Fig. 9. The linear fit to the data gives the slope $dH_{c2}/dT \cong 0.6$ kOe/K. Using the Werthamer-Helfand-Hohenber (WHH) result [18], $H_{c2}(0) = 0.69(dH_{c2}/dT)T_c = 3$ kOe, and the zero-temperature coherence length $\xi(0) = [\Phi_0/2\pi H_{c2}(0)]^{1/2} \cong 30$ nm, where $T_c = 7.3 \pm 0.1$ K is the superconducting transition temperature determined at the peak of $dR_{23}(T)/dT$. The linear Ginzburg-Landau extrapolation gives $H_{c2}(0) = 4.5$ kOe. The obtained values of $H_{c2}(0)$ are much smaller than those reported for Bi so far [12].

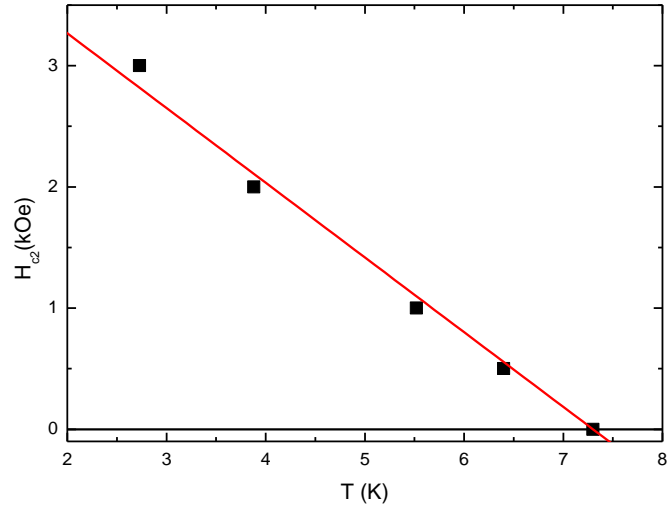

Fig. 9 . The upper critical field $H_{c2}(T)$ obtained from the data of Fig. 8(a). The linear fit gives the slope $dH_{c2}/dT \cong 0.6$ kOe/K.

The superconductivity occurrence at the interface between normal metals and semimetals such as Bi, Sb, and Bi-Sb has been reported long ago by Esaki and Stiles [8]. Instead, in Ref. [9] the superconductivity has been attributed to the formation of Bi-based superconductivity. However, $T_c(0) = 7.3 \pm 0.1$ K (Fig. 8) is much higher than that reported for Bi-Ag phases ($T_c < 3$ K) corroborating the interface-related (Ag/Bi) scenario [8] for the superconductivity. This is also evident from a dramatic sensitivity of the superconductivity to the contact resistance value. As Table 1 illustrates, the sample with higher Ag/Bi contact resistance, $R_c \sim 3$ $\Omega$, possesses no signature for the superconductivity neither in four- nor two-probe configurations. All other samples with lower values of $R_c$ revealed the superconducting signal at least in two-probe measurements.

| Sample | External contact resistance | Internal contact resistance | Superconducting? |
|---|---|---|---|
| 1 | 0.40 $\Omega$ | 0.21 $\Omega$ | Yes |
| 2 | 3.13 $\Omega$ | 0.98 $\Omega$ | No |
| 3 | 0.75 $\Omega$ | 0.97 $\Omega$ | Yes |
| 4 | 0.5 $\Omega$ | 1 $\Omega$ | Yes |
| 5 | 1.05 $\Omega$ | 0.74 $\Omega$ | Yes |
| 6 | 0.46 $\Omega$ | 0.53 $\Omega$ | Yes |

Table 1. Contact resistances $R_c$ for six measured polycrystalline Bi samples. The sample with slightly higher $R_c$ loses the superconductivity. For the definition of internal and external contacts see text and Fig. 7.

The zero-resistance state measured in the four-probe configuration (Fig. 8 a) is of a particular interest because it indicates a percolative system and testify against contamination effects. One can account for the effect as following. The superconducting order parameter $\Psi = \Psi_0 \exp(i\varphi)$ has two components: a magnitude $|\Psi_0| = (n_s)^{1/2}$ and a phase $\varphi$. Phase fluctuations destroy the global phase coherence and hence the superconductivity, although locally the superconducting pairing may exist ($\Psi_0 \neq 0$). Very recently, it has been theoretically demonstrated [19] that a normal metal in a contact with such a phase-fluctuating superconductor increases the phase stiffness and triggers the global superconductivity with the zero-resistance state. It seems, in our case the effect of Ag contacts is two-fold: Ag induces the superconductivity locally, and it may also trigger the global phase coherence.

In conclusion, our results revealed the occurrence of (1) the local superconductivity in powdered Bi samples with $T_c(0) = 8.75 \pm 0.05$ K, and (2) the global superconductivity with $T_c(0) = 7.3 \pm 0.1$ K in polycrystalline Bi triggered by low-resistance Ohmic contacts with silver (Ag) normal metal. The results indicate that the superconductivity emerges at the sample surface. The occurrence of global superconductivity observed for polycrystalline Bi can be accounted for by an enhancement of the superconducting order parameter phase stiffness induced by the normal metal contacts.

This work was supported by FAPESP, CNPq, CAPES and INCT NAMITEC.